\documentclass[fleqn,usenatbib]{mnras}
\usepackage{float}
\usepackage{newtxtext,newtxmath}
\usepackage{afterpage}
\usepackage{placeins}
\usepackage[T1]{fontenc}
\usepackage{xcolor}
\DeclareRobustCommand{\VAN}[3]{#2}
\let\VANthebibliography\thebibliography
\def\thebibliography{\DeclareRobustCommand{\VAN}[3]{##3}\VANthebibliography}
\usepackage{longtable}
\usepackage{supertabular}

\usepackage{orcidlink}

\usepackage{graphicx}	
\usepackage{amsmath}	

\title[Quasi-periodic oscillation and broadband noise characteristics of RX J0440.9+4431]{Broadband noise and quasi-periodic oscillation characteristics of the X-ray pulsar RX J0440.9+4431}

\author[P. P. Li et al.]{
P. P. Li,$^{1,2}$ L. Tao~\orcidlink{0000-0002-2705-4338},$^{1}$\thanks{E-mail: taolian@ihep.ac.cn} R. C. Ma,$^{1, 2}$ M. Y. Ge~\orcidlink{0000-0002-2749-6638},$^{1}$ Q. C. Zhao,$^{1, 2}$ S. J. Zhao,$^{1, 2}$ L. Zhang,$^{1}$ Q. C. Bu~\orcidlink{0000-0001-5238-3988},$^{3}$ \newauthor{L. D. Kong~\orcidlink{0000-0003-3188-9079},$^{3}$ Y. L. Tuo~\orcidlink{0000-0003-3127-0110},$^{3}$   L. Ji~\orcidlink{0000-0001-9599-7285},$^{4}$ S. Zhang,$^{1}$ J. L. Qu,$^{1}$ S. N. Zhang~\orcidlink{0000-0001-5586-1017},$^{1}$ Y. Huang,$^{1}$ X. Ma,$^{1}$}  \newauthor{W. T. Ye,$^{1, 2}$ Q. C. Shui,$^{1, 2}$ }
\\
$^{1}$Key Laboratory of Particle Astrophysics, Institute of High Energy Physics, Chinese Academy of Sciences, 100049 Beijing, People’s Republic of China\\
$^{2}$Uinversity of Chinese Academy of Sciences, Chinese Academy of Sciences, 100049 Beijing, People’s Republic of China\\
$^{3}$Institut f\"ur Astronomie und Astrophysik, Kepler Center for Astro and Particle Physics, Eberhard Karls Universit\"at, Sand 1, D-72076 T\"ubingen, Germany\\
$^{4}$School of Physics and Astronomy, Sun Yat-sen University, Zhuhai, 519082, People’s Republic of China}

\date{Accepted XXX. Received YYY; in original form ZZZ}

\pubyear{2015}

\begin{document}
\label{firstpage}
\pagerange{\pageref{firstpage}--\pageref{lastpage}}
\maketitle

\begin{abstract}
We present a comprehensive timing analysis on the Be/X-ray binary pulsar RX J0440.9+4431 using observations from \textit{NICER} and \textit{Insight}-HXMT during the 2022--2023 outburst. The power density spectrum (PDS) of RX J0440.9+4431 exhibits typical aperiodic variability in X-ray flux across a wide frequency range. During a super-critical accretion state, we detect quasi-periodic oscillations (QPOs) at 0.2--0.5\,Hz in the light curves of five pulses for RX J0440.9+4431. The observed QPOs manifest during flares, while the flares appear at the peaks of the pulse profiles on a timescale of seconds and are primarily caused by an increase in hard photons. These flares can be explained by increased material ingestion in the accretion column at a fixed phase, primarily generating hard photons. Alternatively, an increase in accretion rate, independent of phase, may result in highly beamed hard photons within the accretion column, causing the flares. We argue the origin of QPOs to instabilities within the accretion flow. Additionally, we find that the break frequencies in the noise power spectra align well with $\propto L_{\mathrm{x}}^{3 / 7}$ across three orders of magnitude in the luminosity, which points to a relatively strong magnetic field in RX J0440.9+4431, estimated to be \textasciitilde$10^{13}$\,G.


\end{abstract}




\section{Introduction}
In the analysis of X-ray pulsars (XRPs), the power density spectrum (PDS) serves as a commonly used tool for scrutinizing light curves. It presents the power distribution of signals in the frequency domain, thereby unveiling potential signal structures in observational data \citep[e.g.][]{2014Tian,2016Wang}. Within the PDS of XRPs, in addition to naturally displaying inherent periodic peaks corresponding to pulse components and their harmonics, there is also a strong aperiodic variability in X-ray flux over a wide frequency range, characterized by band-limited noise and possible quasi-periodic oscillations \citep[QPOs; e.g.][]{2008Reig,2010James}. The power of band-limited noise increases at lower frequencies and exhibits a flattening towards higher frequencies, approximated by a broken (or double-broken) power law. This phenomenon is similar to what is observed in accreting black holes \citep[BHs; e.g.][]{2015Stiele} and active galactic nuclei \citep[AGN; e.g.][]{2004McHardy}.

The aperiodic noise can be naturally explained by the propagating fluctuations model \citep{1997Lyubarskii,2001Churazov}. According to this model, the initial fluctuations in mass accretion rate occur at different radial coordinates in the accretion disc, and then propagate towards the central object, causing local variations in mass accretion rate at all radii. The variability of the local mass accretion rate leads to photon flux variability observed from different parts of the accretion disc. The time-scales of the initial fluctuations are comparable to or smaller than the local Keplerian time-scale, while the process of viscous diffusion in the disc effectively suppresses variability above the frequency of the initial fluctuations \citep{2018Mushtukovp,2019Mushtukov}. Therefore, the observed break frequency in the PDS is closely related to the inner disc radius \citep{2009Revnivtsev}. As the X-ray luminosity varies, any corresponding changes in the size of the magnetosphere will result in a corresponding alteration in the break frequency \citep{2009Revnivtsev,2020Doroshenko}. Even though photons mostly originate from the neutron star (NS) surface, the time-scales of photon diffusion from a hotspot/accretion column and reprocessing of X-ray flux by the atmosphere are much shorter than the Keplerian time-scale at the inner disc radius. Therefore, it is believed that the observed aperiodic variability of X-ray flux replicate fluctuations in mass accretion rate at the inner disc radius \citep{2019Mushtukov}. Consequently, for known magnetic field strength in XRPs, we can use the measured break frequency in the PDS to estimate the inner radius of the disc \citep{2016Bodaghee}. For XRPs with unknown magnetic field, we can estimate it using the relationship between break frequency and luminosity \citep{2014Doroshenko,2022MM}.

QPOs have been detected in multiple highly magnetized NS systems, only concentrated in the low-frequency range \citep{2011Devasia}, from \textasciitilde10\,mHz to \textasciitilde1\,Hz, referred to as mHz QPOs. These QPOs are generally thought to be related to the rotation of the inner accretion disc. Two widely used models for explaining the mHz QPO frequency ($\nu_{\rm qpo}$) are the Keplerian Frequency Model \citep[KFM;][]{1987van} and the Beat-Frequency Model \citep[BFM;][]{1985Alpar}. In the KFM, QPOs are generated because some inhomogeneous structures in the Keplerian disc attenuates the pulsar beam regularly. Therefore, the Keplerian frequency is $\nu_{\rm k} = \nu_{\rm qpo}$. However, the KFM is valid only when the spin frequency is lower than the QPO frequency, making it unsuitable for sources like 4U 0115+63 \citep{2012Li}, 4U 1901+03 \citep{2011Jamesb}, and Cen X--3 \citep{2008Raichur}. The BFM, on the other hand, assumes that QPOs result from the accretion process, controlled by the interaction (matching) between the magnetosphere co-rotating with the star and matter at the unstable inner accretion disc, then the luminosity will display a 'beat' frequency ($\nu_{\rm qpo} = \nu_{\rm k} - \nu_{\rm spin}$). This model has been employed to explain QPOs in sources like EXO 2030+375 \citep{1989Angelini} and 4U 1901+03 \citep{2011James}.


However, there remain certain properties of mHz QPOs that elude explanation by these models \citep{2005Qu,2022Ma}. Furthermore, due to limitations imposed by observational instruments, there are only a few analyses of mHz QPOs across a broader range of luminosities and energies. Hence, a comprehensive exploration of mHz QPO properties holds promise for advancing theoretical research in the future.

The X-ray source RX J0440.9+4431 contains a rotating NS ($P_{\rm spin}$ = 205\,s) in orbit around a Be primary \citep{1997Motch}. The source experienced its brightest-ever recorded outburst at the end of 2022 \citep{2022Nakajima,2023ATelpal,2023ATelColey}, reaching a peak flux of \textasciitilde$2.25$\,Crab, allowing us to study this source across a wider range of luminosities. Through the investigation of the evolution of pulse profiles and spectra during this outburst, the critical luminosity of the source was determined to be \textasciitilde$3\times10^{37}\ {\rm erg\ \rm s^{-1}}$ \citep{2023Salganik,2023Mandal,2023lipp}. Additionally, \cite{2023Doroshenko} presented the first polarization results for RX J0440.9+4431, deriving constraints on the pulsar geometry. Recently, \cite{2023Malacaria} reported the discovery of \textasciitilde0.2\,Hz QPOs from this source observed with \textit{Fermi}-GBM. This marks the first report of QPOs associated with RX J0440.9+4431. However, the sensitivity of \textit{Fermi}-GBM and the limited availability of time-tagged event data have constrained further investigation into this QPO. Leveraging the advantages of the broad energy range and large effective area offered by the Hard X-ray Modulation Telescope \citep[\textit{Insight}-HXMT;][]{zhang2020} and \textit{Neutron star Interior Composition Explorer} \citep[\textit{NICER};][]{2016SPIE}, we employ dynamic windows to conduct PDS analysis, ultimately discovering QPOs in five pulses. Moreover, we discuss the magnetic field of RX J0440.9+4431 using the relationship between break frequency and luminosity. The structure of this article is as follows. Section~\ref{sec:2} describes the observations and data reduction. The results are presented in Section~\ref{sec:3}. Discussion and conclusion are summarized in the Section~\ref{sec:4} and Section~\ref{sec:5}, respectively.

\begin{figure}
    \includegraphics[width=\columnwidth]{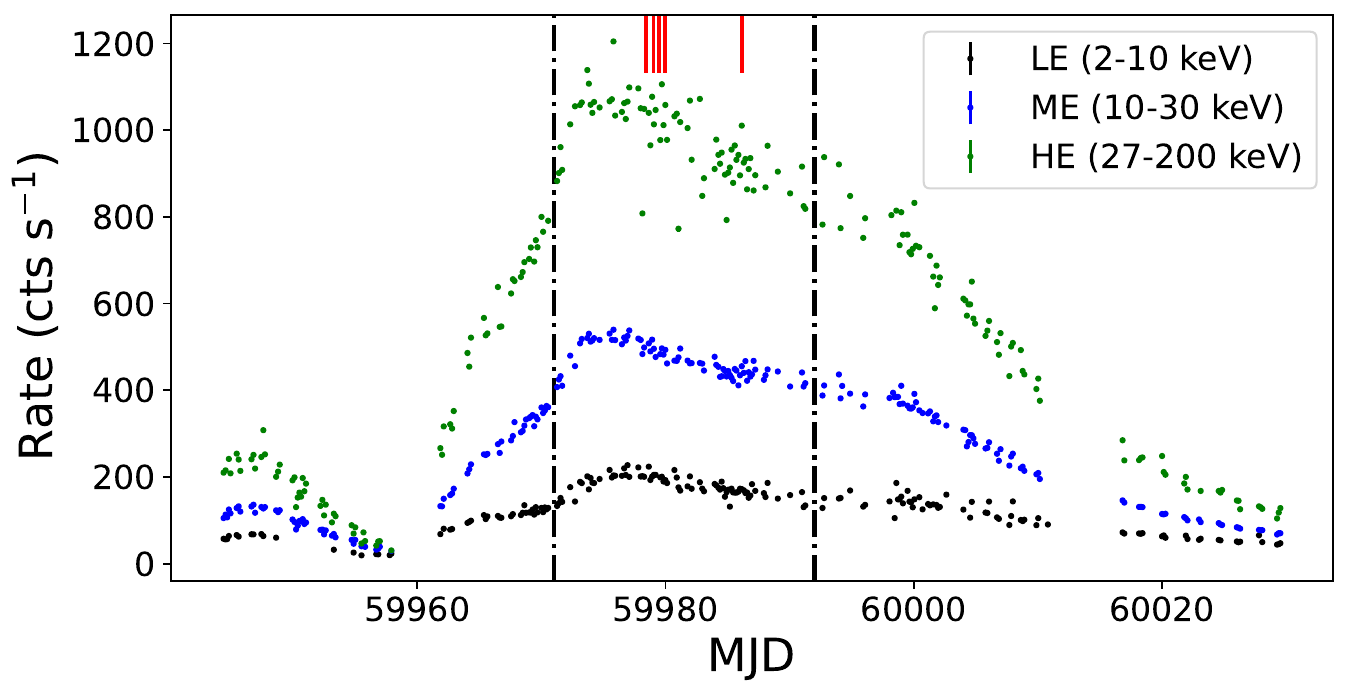}
	\includegraphics[width=\columnwidth]{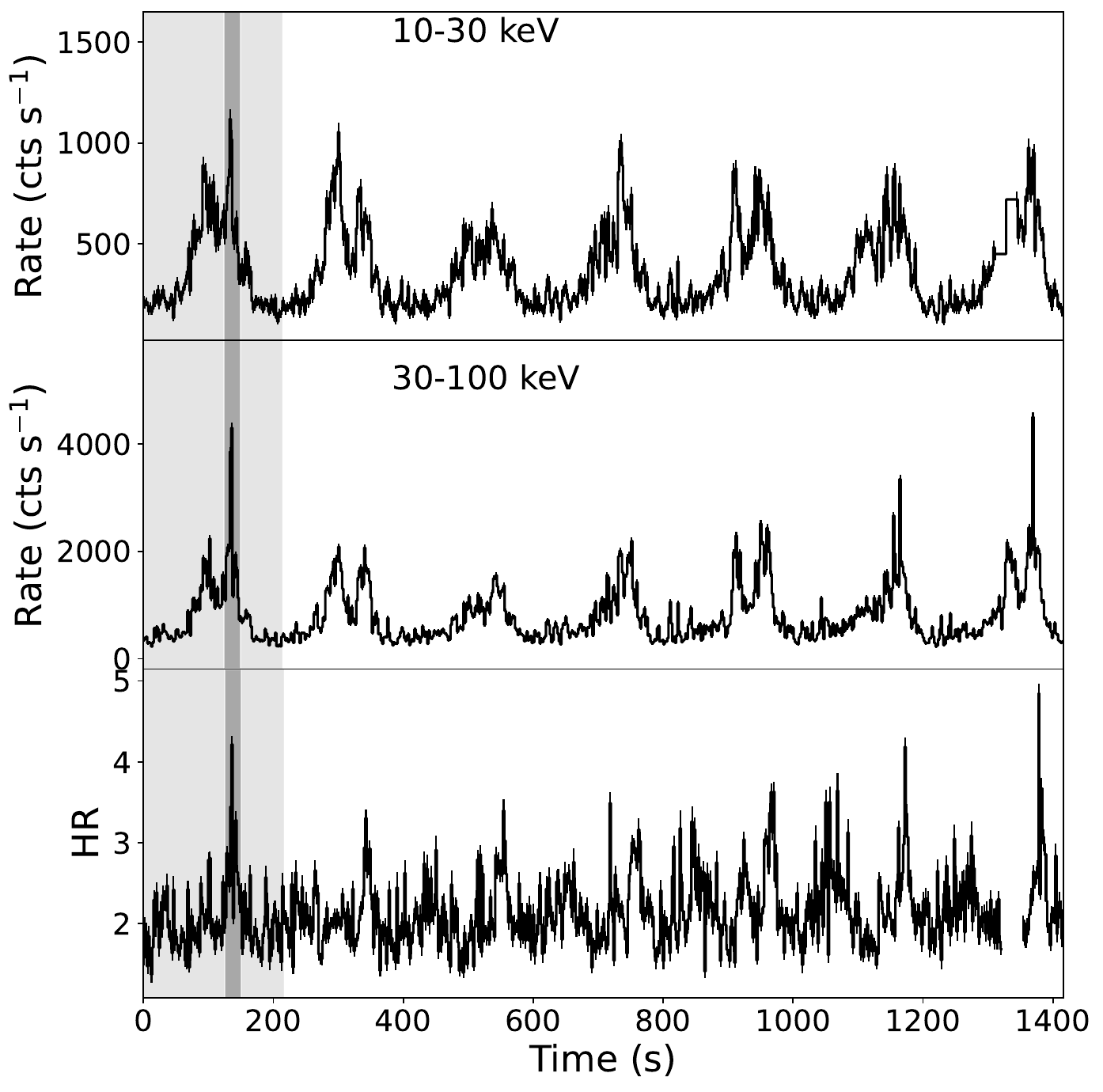}
    \caption{Upper panel: Light curve of RX J0440.9+4431
obtained with \textit{Insight}-HXMT during the 2022--2023 outburst \citep{2023lipp}. Between the dashed lines (MJD 59971; MJD 59992), the source is in a supercritical accretion state, occasionally flaring at the pulse profile peaks. The red vertical lines mark the times of the QPO detection, which will be detailed in the following sections. Lower panel: Light curve with a bin size of 2\,s and hardness ratio (30--100\,keV/10--30\,keV) obtained from \textit{Insight}-HXMT around MJD 59978.44. The gray shadow represents one cycle of the pulse profile. The flares marked by the dark gray shaded areas exhibit QPO.}
  \label{fig:1}
\end{figure}

\begin{figure}
	\includegraphics[width=0.9\columnwidth]{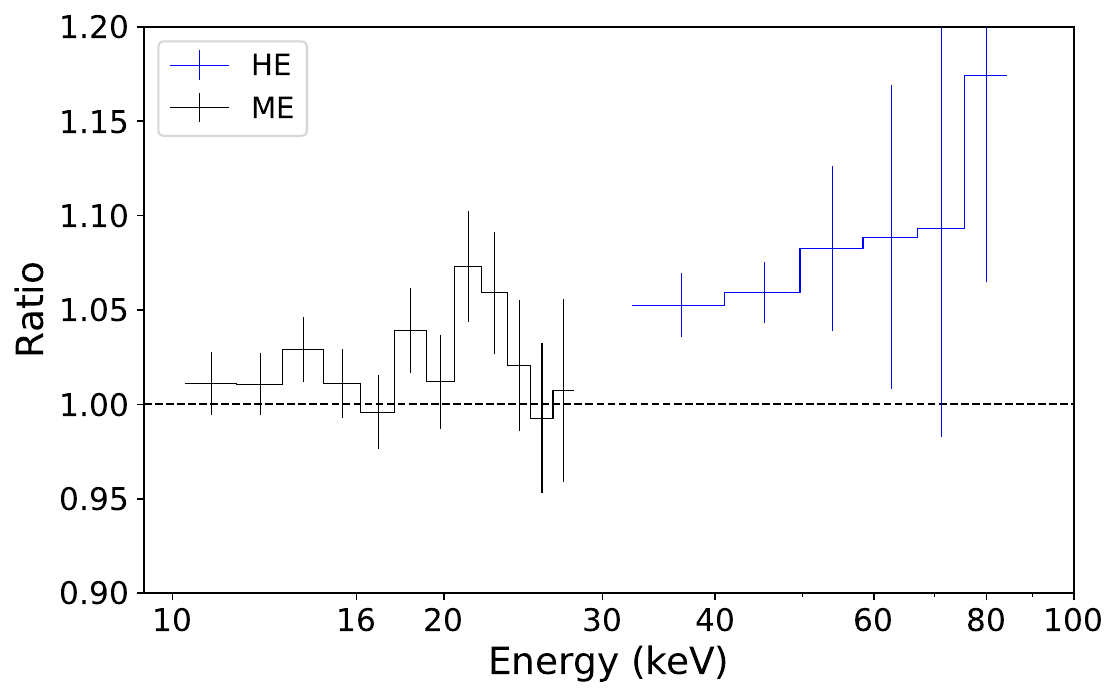}
    \caption{Flux ratios between the pulses with flares (the shaded region in the lower panel of Fig.~\ref{fig:1}) and adjacent pulses without flares.}
  \label{fig:5}
\end{figure}


\begin{figure*}
	\includegraphics[width=0.75\textwidth]{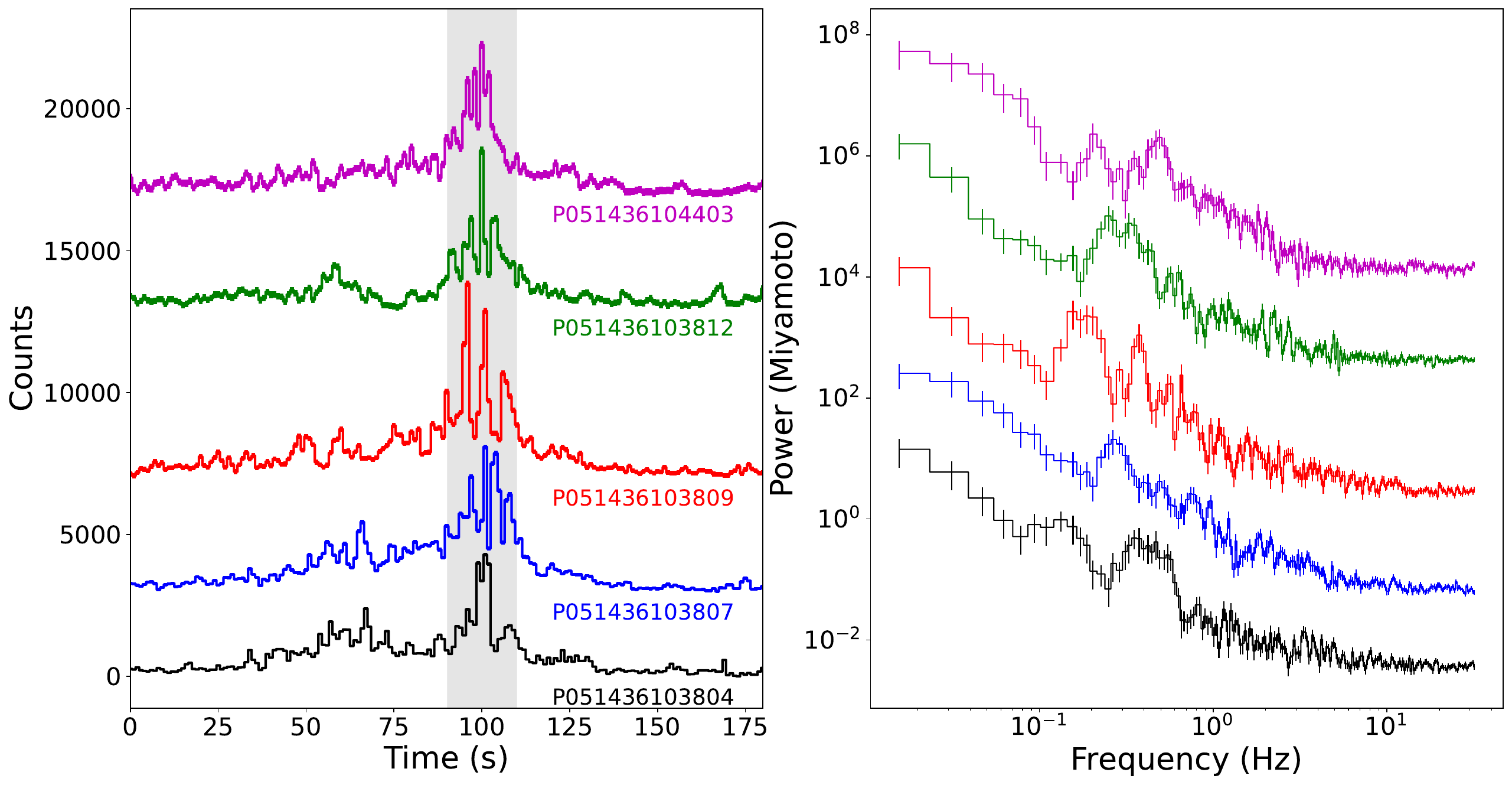}
    \caption{Net light curve with QPOs in 30--100\,keV energy band, each represented by a distinct color (left panel; bin size 1\,s) and the corresponding PDS (right panel). Counts and power are plotted with certain offsets for clarity. QPOs manifest within the flares at the peaks of the pulse profile, marked by gray shading.}
  \label{fig:2}
\end{figure*}


\begin{figure}
	\includegraphics[width=0.9\columnwidth]{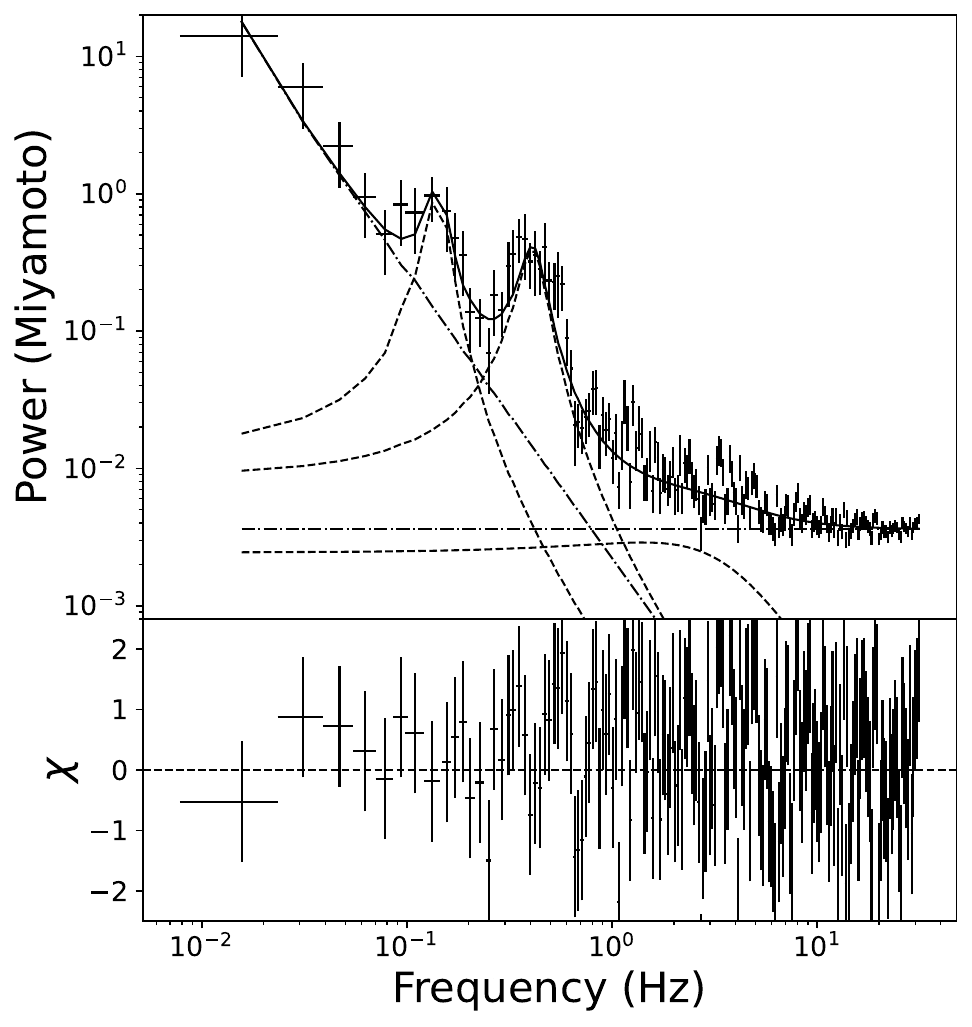}
    \caption{Miyamoto normalized power density spectrum of ObsID P051436103804 and the residuals from the PDS fitting. The dashed line represents the Lorentzian model, and the dashed dotted line represents the power-law model for fitting.}
  \label{fig:3}
\end{figure}

\begin{figure}
    \centering
	\includegraphics[width=0.9\columnwidth]{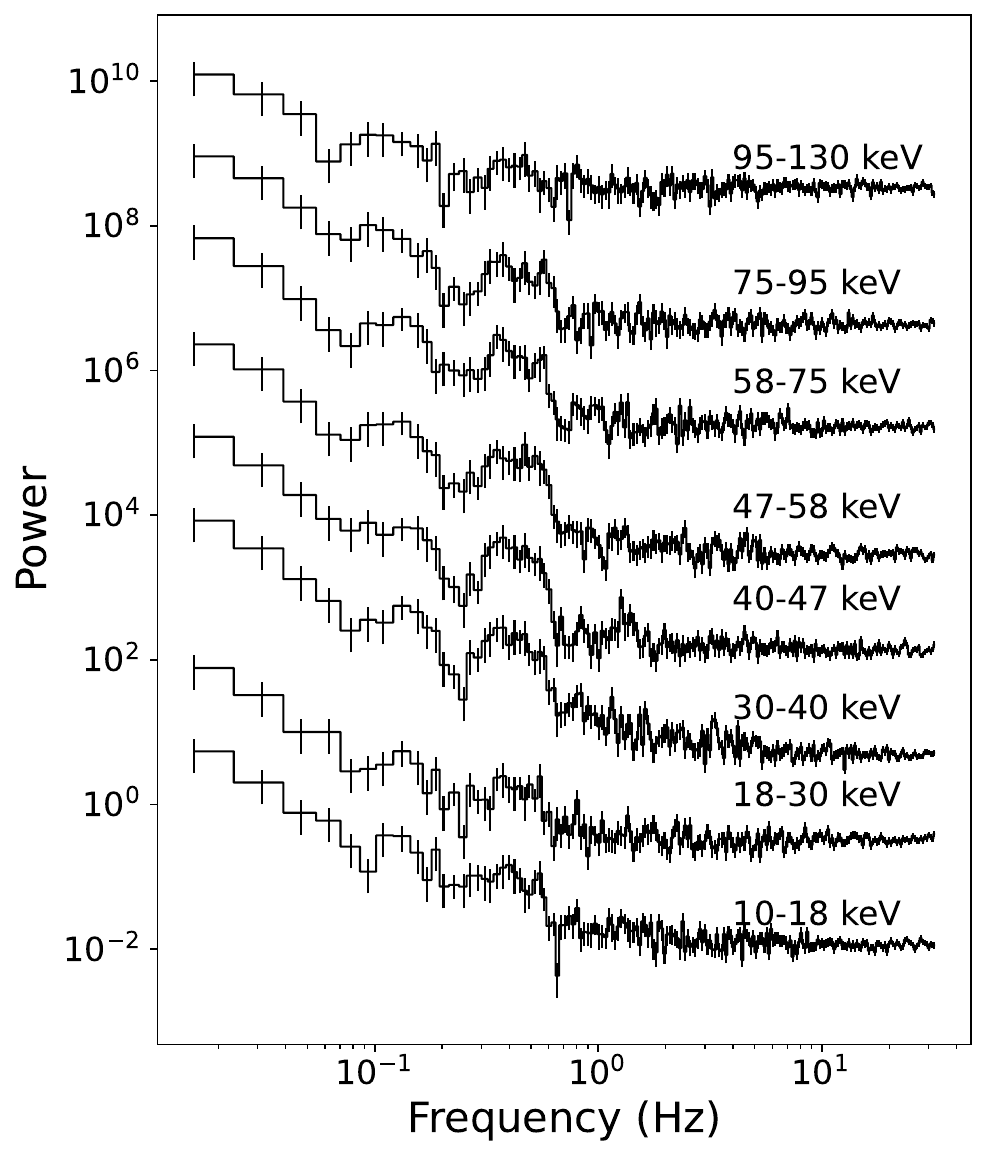}
    \includegraphics[width=0.85\columnwidth]{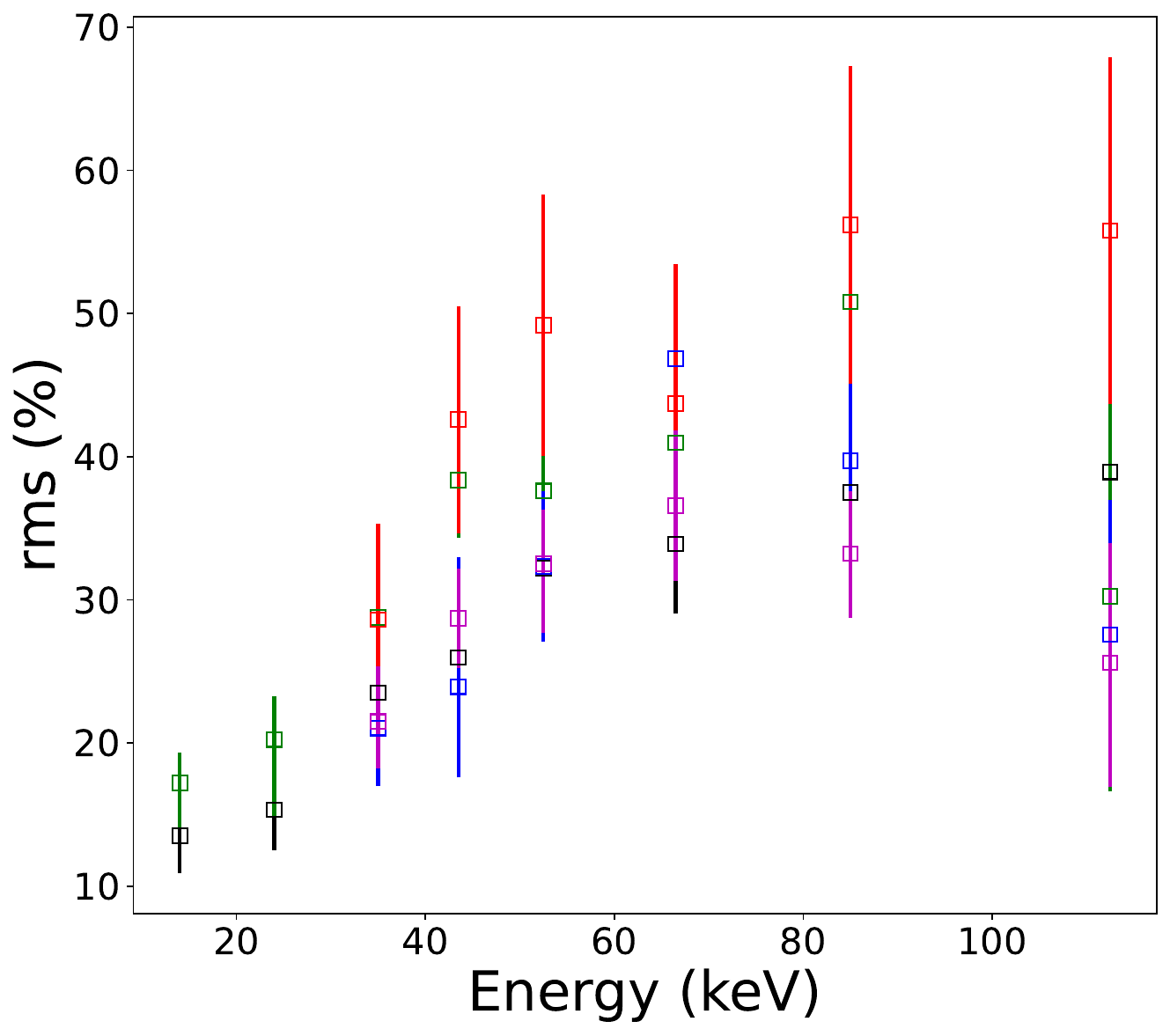}
    \caption{Upper panel: PDS in different energy ranges for ObsID P051436103804, as a representative, with specific offsets applied for plotting clarity. Lower panel: Dependence of QPO rms on energy for all QPOs. The colors correspond to those used in Fig.~\ref{fig:2}}
  \label{fig:4}
\end{figure}


\section{Observations and Data Reduction}
\label{sec:2}
\textit{Insight}-HXMT, stands as Chinese first X-ray astronomy satellite, offering exceptional capabilities across a wide energy spectrum (1--250\,keV) with a substantial effective area in the hard X-ray range \citep{zhang2020}. It comprises three collimated telescopes: the high-energy X-ray telescope \citep[HE, 20--250\,keV, 5100\,$\rm cm^2$,][]{2020SCPMALiu}, the medium-energy X-ray telescope \citep[ME, 5--30\,keV, 952\,$\rm cm^2$,][]{2020SCPMACao}, and the low-energy X-ray telescope  \citep[LE, 1--15\,keV, 384\,$\rm cm^2$,][]{2020SCPMAChen}. These telescopes enable observations with time resolutions of 1\,ms, 276\,$\upmu$s, and\,25\,$\upmu$s, respectively.
\textit{Insight}-HXMT conducted intensive observations on RX J0440.9+4431 from MJD 59944 to 60029, and the ObsIDs are listed in \cite{2023lipp}. Details regarding data processing procedures and criteria for selecting good time intervals (GTIs) can be found in \cite{2023lipp}. We performed PDS analysis on all observations to search for potential QPOs.

\textit{NICER} has been stationed on the International Space Station and operated by NASA since June 2017 \citep{2016SPIE}. Its primary instrument, the X-ray Timing Instrument (XTI), consists of 56 X-ray concentrator optics (XRC) and silicon drift detector (SDD) pairs. It is capable of soft X-ray spectroscopy in the range of 0.2 to 12\,keV with a timing precision of $<300$\,ns and a effective area of \textasciitilde1900\,$\rm cm^2$ at 1.5\,keV. This complements the low-energy detection capabilities of \textit{Insight}-HXMT. We select the observational data of RX J0440.9+4431 obtained by \textit{NICER} during the period from MJD 59942 to MJD 60046 with long  exposure time. The data processing procedures are conducted following the same methodology as described by \cite{2023lipp}.

\begin{table*}
    \centering
    \caption{Lorentzians fitting parameters for the narrow peaks in the PDS from \textit{Insight}-HXMT/HE. $\nu$, rms, Q, and Sig represent the centroid frequency, fractional rms amplitude, quality factor, and significance, respectively. Subscript numbers indicate different Lorentzian components. Our primary focus is on QPOs with a significance $>3\sigma$ and clear modulation in the light curve.} 
    \renewcommand\arraystretch{1.37}
    \label{tab:1}
    \begin{tabular}{|c|c|c|c|c|c|c|c|c|c|c|c|c|}
    \hline
        ObsIDs  & Time (MJD) & $\nu$$_{0}$ (Hz)  & rms$_{0}$ (\%)  & Q$_{0}$  & Sig$_{0}$ ($\sigma$)  & $\nu$$_{1}$ (Hz) & rms$_{1}$ (\%)  & Q$_{1}$  & Sig$_{1}$ ($\sigma$)  \\ \hline
        P051436103804  & 59978.44 & $0.14^{+0.02}_{-0.03}$ & $23.51^{+9.23}_{-6.85}$ & $3.99^{+5.04}_{-3.94}$ & 1.72& $0.41^{+0.04}_{-0.03}$ & $27.18^{+3.55}_{-4.03}$ & $3.26^{+1.04}_{-0.91}$  & 3.33   \\ 
        P051436103807  & 59979.03 & $0.29^{+0.02}_{-0.01}$ & $25.20^{+4.62}_{-4.58}$ & $4.23^{+4.34}_{-1.76}$ & 3.94 & ~.~.~. & ~.~.~. & ~.~.~. & ~.~.~.  \\ 
        P051436103809  & 59979.47 & $0.18^{+0.02}_{-0.02}$ & $33.11^{+8.15}_{-8.08}$ & $8.04^{+6.72}_{-5.59}$ & 2.13 & $0.37^{+0.01}_{-0.01}$ & $24.93^{+5.12}_{-5.07}$ & $14.35^{+11.05}_{-6.43}$ & 3.22   \\ 
        P051436103812  & 59979.94 & $0.32^{+0.02}_{-0.02}$ & $33.96^{+3.34}_{-4.25}$ & $3.23^{+1.35}_{-0.85}$ & 4.11 & ~.~.~. & ~.~.~. & ~.~.~. & ~.~.~.  \\ 
        P051436104403  & 59986.14 & $0.21^{+0.02}_{-0.01}$ & $15.41^{+4.51}_{-4.58}$ & $10.57^{+22.27}_{-10.56}$ & 1.75 & $0.48^{+0.02}_{-0.01}$ & $26.76^{+3.29}_{-3.36}$ & $5.46^{+2.81}_{-1.84}$ & 4.04    \\ 
     \hline
    \end{tabular}
\end{table*}


\begin{figure}
	\includegraphics[width=0.9\columnwidth]{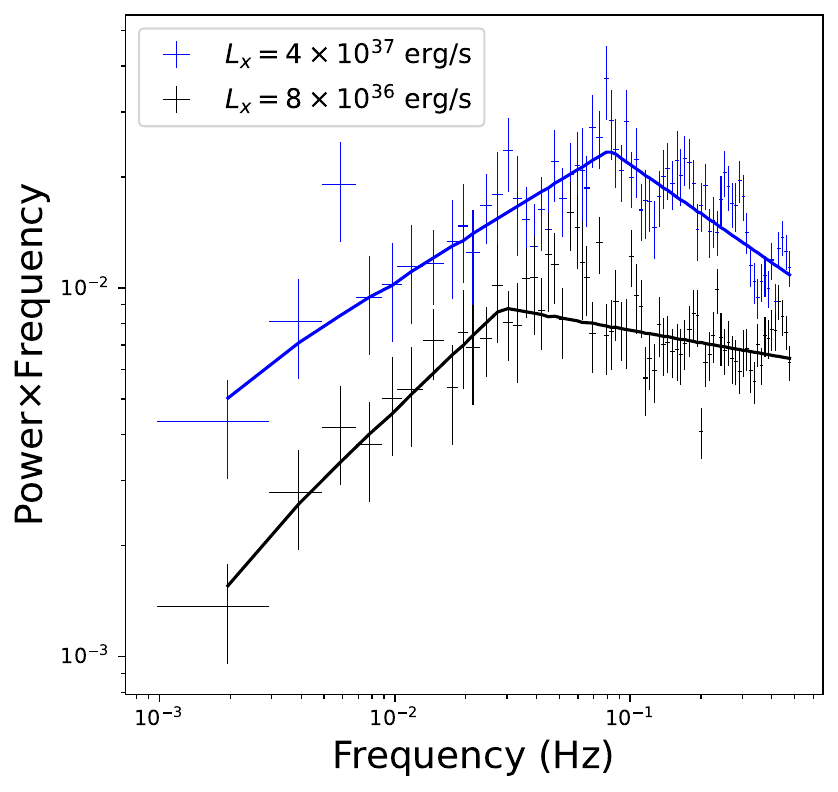}
    \caption{Examples of representative power spectra at different luminosities. The solid lines show the best-fitting models.}
  \label{fig:6}
\end{figure}

\begin{figure}
	\includegraphics[width=0.9\columnwidth]{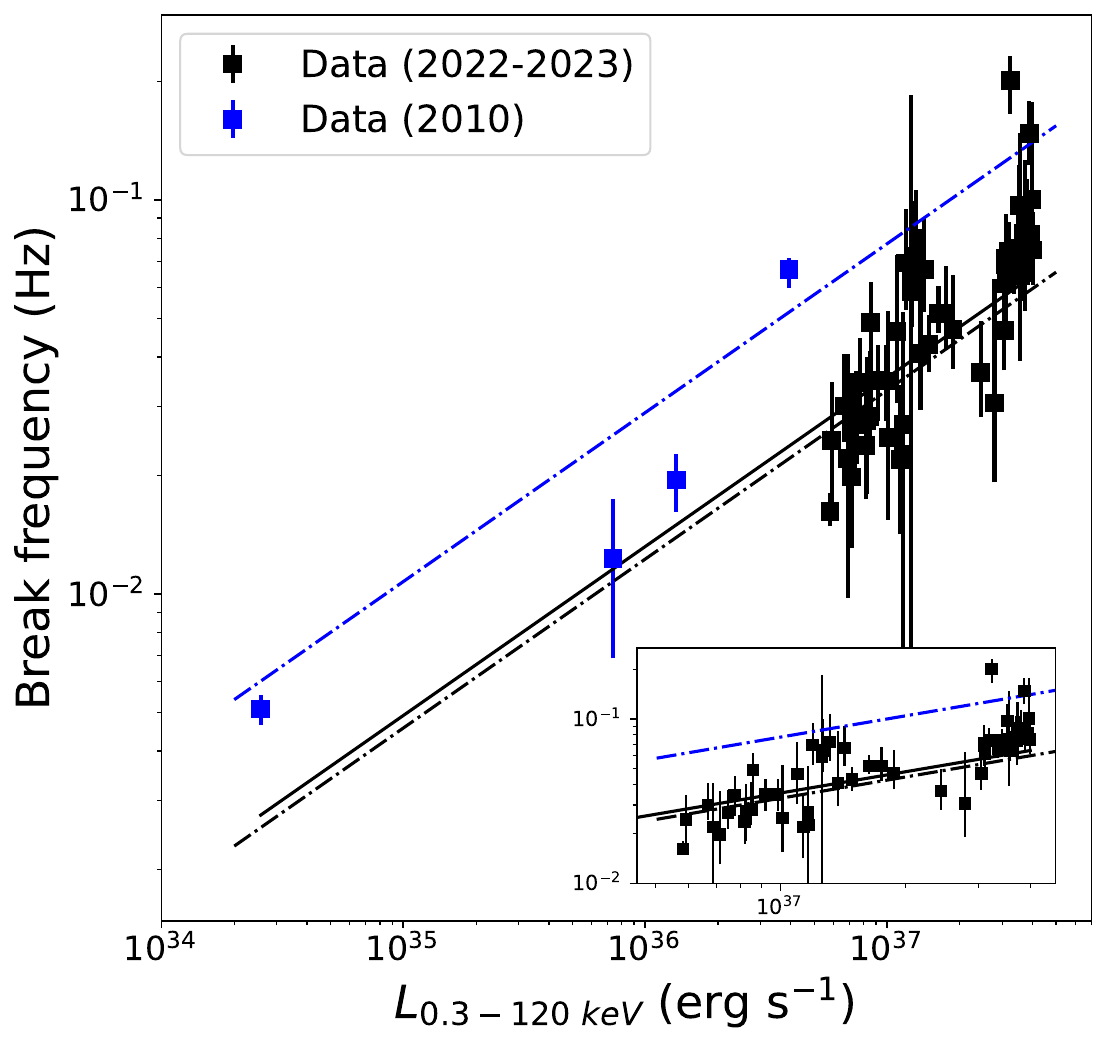}
    \caption{Relationship between the break frequencies of the noise power spectra and the 0.3--120\,keV luminosities. The black points represent the results using the 0.5--10\,keV \textit{NICER} light curve in this study, and the black dashed dotted line indicates the modeled results. The blue points and blue dashed dotted line are from previous studies \citep{2012Tsygankov}. The solid black line shows the model fitting in combination of black and blue points. The insert zooms in on the data at high luminosities.}
  \label{fig:7}
\end{figure}


\section{Analysis and Results}
\label{sec:3}
\subsection{Quasi-periodic oscillation}
\label{sec:3.1}
Recently, \cite{2023Malacaria} reported that \textit{Fermi}-GBM triggered on RX J0440.9+4431 several times during the brightest period of the outburst (MJD 59970--60000). All GBM triggers occur at the peak of the sine-like pulse profile and show one or more flares. A QPO at $\sim$0.2\,Hz was discovered at the pulse peak, suggesting a potential correlation between the QPO and these flares. Inspired by their work, we also carefully check the light curves of RX J0440.9+4431 observed by the \textit{Insight}-HXMT. When the pulse profile of RX J0440.9+4431 transits from the double-peak structure to a single-peak structure \citep{2023lipp}, and the main peak splits into two wings, i.e., in the supercritical state, a significant increase in photon counts (exhibiting as a flare) is occasionally observed in the right wing, as shown in Fig.~\ref{fig:1}. Moreover, in the 30--100\,keV energy band, the count rate of the right wing with flares is more than twice that of those without flares, and the hardness ratio (HR) is also larger. We further plot the flux ratios between the periods with flares and adjacent periods without flares in Fig.~\ref{fig:5}, which more clearly illustrates that flares are primarily associated with an increase in high-energy photons.

Considering that the variations in the right wing may be related to the hard component, we choose to calculate the PDS using the 30--100\,keV energy range of \textit{Insight}-HXMT/HE net light curve in order to search for potential QPO signals. Initially, we average the PDS for each exposure using the {\tt Stingray} package in {\tt Python} \citep{matteo2022}, but no suspected QPOs are observed. Therefore, we decide to use a dynamic window for PDS analysis: averaging power spectra for each 320\,s window using 64\,s segments and 0.015625\,s time bins, corresponding to a frequency resolution of 1/64\,Hz and a Nyquist frequency of 32\,Hz. This window is then moved every 25\,s. Ultimately, we have identified QPOs in five pulses, with the corresponding PDS provided in Fig.~\ref{fig:2}. Upon further examination of the PDS in the windows covering these pulses, it is revealed that the QPOs only appear when the right wing of the pulse profile is covered by the window. In other words, these QPOs originate from short-scale flares on the pulse peaks, consistent with the results reported by \cite{2023Malacaria}. Although we analyze \textit{NICER} data using the same method, no QPO signals are found.



To characterize quasi-periodic variability, we carry out the PDS fitting using the {\tt XSPEC} fitting package. The "spectrum" file and diagonal response are created using {\tt flx2xsp} \citep{2012Ingram}. Following \cite{2023Malacaria}, our PDS model includes a power-law continuum, several Lorentzian features \citep{2002Belloni}, and a constant to parametrize Poisson noise (see Fig.~\ref{fig:3}).
The power ($P(\nu)$) for a Lorentzian component can be formulated as
\begin{equation}
P(\nu)=\frac{r^2 \Delta}{\pi} \frac{1}{\Delta^2+\left(\nu-\nu_0\right)^2},
\end{equation}
where $\nu_0$ is the centroid frequency, $\Delta$ represents the half width at half maximum (HWHM), and $r$ is the integrated fractional rms under the Lorentzian. The quality factor $Q$ is defined as $\nu_{0}$/2$\Delta$, providing a measure of the relative width of the Lorentzian. Typically, components with $Q>2$ are identified as QPO. The fitting results are presented in Table~\ref{tab:1}. Our primary focus is on QPOs with a significance $>3\sigma$ and clear modulation in the light curve. The centroid frequency is in the range of 0.2--0.5\,Hz. The frequency appears random and is independent to the source intensity.

We further calculate the energy dependence of QPO features, by creating PDS for different energy bands (30--40, 40--47, 47--58, 58--75, 75--95, and 95--130 keV) using HE. For ObsIDs P051436103804 and P051436103812, where the QPO was detected, simultaneous ME observation data is available, thus we also analyze two additional energy bands: 10--18\,keV and 18--30\,keV. For other ObsIDs, it is observed that the pulsation containing QPO does not fall within the GTIs for ME observations. Furthermore, these five observations containing QPOs also lack simultaneous observations from either \textit{NICER} or LE, preventing us from analyzing the PDS in lower energy bands. The upper panel of Fig.~\ref{fig:4}, using ObsID P051436103804 as an example, illustrates the power spectra in different energy bands. The energy dependence of fractional rms amplitude for all QPOs are presented in the lower panel of Fig.~\ref{fig:4}. The rms increases from \textasciitilde15\% at 10--18\,keV to \textasciitilde40\% at 58--75\,keV. Although the rms values remain relatively high above 75\,keV, the errors are also substantial. The lack of clear high-energy QPOs may be attributed to the lower count rate. 

\subsection{Broadband noise}
\label{sec:3.2}

The variations in X-ray flux in XRPs are determined by fluctuations in the mass accretion rate at the inner edge of the accretion disc. We can relate the observed break frequency to the mass accretion rate with
\begin{equation}
2 \pi v_k=(G M)^{1 / 2} R_{\mathrm{m}}^{-3 / 2},
\end{equation}
where 
\begin{equation}
\begin{aligned}
R_{\mathrm{m}} & =2.7 \times 10^8 \Lambda \mu_{30}^{4 / 7} \dot{M}_{17}^{-2 / 7} m^{-1 / 7} \mathrm{~cm} \\
& =1.8 \times 10^8 \Lambda B_{12}^{4 / 7} \dot{M}_{17}^{-2 / 7} m^{-1 / 7} R_6^{12 / 7} \mathrm{~cm}
\end{aligned}
\end{equation}
 is the inner disc radius \citep{2002Frank,2022Mushtukov}. Combining equations (2) and (3), we get
\begin{equation}
v_k=0.76 \Lambda^{-3 / 2} m^{\frac{5}{7}} R_6^{-\frac{18}{7}} B_{12}^{-\frac{6}{7}} \dot{M}_{17}^{\frac{3}{7}},
\end{equation}
where $B_{12}$ is the magnetic field strength $B$ at the NS surface in units of $10^{12}$\,G, $\dot{M}_{17}$ is mass accretion rate in units of $10^{17}\rm g\rm~s^{-1}$, $m$ is the NS mass in units of solar masses $M_{\rm \odot}$ and $R_6$ is the
NS radius in units of $10^{6}$ cm. $\Lambda = 0.5$ being a commonly used value for the case of accretion through the disc \citep{1979Ghosh,2014Lai}. Assuming that the break frequency equals or is a fraction of the Keplerian frequency \citep{2004King,2019Mushtukov}, and all the kinetic energy of infalling matter is converted to radiation at the stellar surface ($L_{\mathrm{x}} = G M \dot{M}/R$), the magnetic field can be inferred.


\cite{2012Tsygankov} previously analyzed the relationship between the break frequency in the noise power spectrum and the bolometric luminosity of RX J0440.9+4431; however, due to the faintness of the source and the short duration of the observations, the data points are very limited. Here, we analyze the aperiodic noise in a broader luminosity range from \textit{NICER} data. The same approach as \cite{2009Revnivtsev} is adopted to suppress pulsations. Considering the length of GTIs and the spin period of the source, we set the segment length to 512\,s to generate individual power spectra. We select observations with sufficient exposure, and average three to five individual power spectra together. This ensures a good signal-to-noise ratio and better removal of pulse components in the PDS. \textit{Insight} - HXMT data is not used as individual GTIs mostly do not exceed 500\,s, making it difficult to find ObsIDs that meet the requirement of having three segments of 512 seconds in light curve. Additionally, if the interval between the first and third segments is too long, it becomes challenging to effectively remove pulse components. 

The resulting spectrum is described by a broken power-law ({\tt bknpower} in {\tt XSPEC}), as shown in Fig.~\ref{fig:6}. The dependence of the break frequencies on the source luminosity is presented in Fig.~\ref{fig:7}. The luminosity is derived from the work of \cite{2023lipp}, with the luminosity for each data point calculated through interpolation based on adjacent time observations. The data points from \cite{2012Tsygankov} are also included (blue points in Fig.~\ref{fig:7}). By noticing that the values from \cite{2012Tsygankov} were obtained in the 3--100\,keV band with a distance of 3.3\,kpc \citep{2005Reig}, we extrapolate the flux to the 0.3--120\,keV band using the spectral parameters from Table 1 of \cite{2012Tsygankov}, and calculate the luminosities with an updated distance of 2.4\,kpc \citep{2021Bailer-Jones}. As shown in Fig.~\ref{fig:7}, over a broad luminosity range, the break frequencies appear well in agreement with $\propto L_{\mathrm{x}}^{3 / 7}$.


\section{Discussion}
\label{sec:4}
\subsection{QPO generation}
RX J0440.9+4431 shows a single-peaked pulse profile splitting into two wings when the luminosity exceeds \textasciitilde$3\times10^{37}\ {\rm erg\ \rm s^{-1}}$ \citep{2023lipp,2023Salganik,2023Mandal}. The right wing occasionally generates flares with a timescale of seconds, primarily caused by an increase in hard photons, as shown in Fig.~\ref{fig:1}. Similar flares also triggered \textit{Fermi}-GBM, with a \textasciitilde0.2\,Hz QPO discovered in the averaged periodogram of all trigger data \citep{2023Malacaria}. This QPO only appears at the peak of its sine-like pulse profile. The large effective area of \textit{Insight}-HXMT allows us to calculate an average PDS using 320\,s bin and search for QPO in a smaller time windows. Five QPOs are finally found, located in five pulses separately (see Fig.~\ref{fig:2}). Except for the pulse displaying QPO in ObsID P051436103809, which corresponds to the second trigger listed in \cite{2023Malacaria}, the remaining pulses were not observed by \textit{Femi}-GBM. Besides, the rest of the triggers in Table 1 of \cite{2023Malacaria} are not included in the GTIs of \textit{Insight}-HXMT observations. The discovered QPOs are also manifested as large modulation in the flares (see Fig.~\ref{fig:2}), consistent with the findings reported by \cite{2023Malacaria}. It is important to emphasize that not all flares at the right wing of pulse profiles exhibit quasi-periodic modulation. Furthermore, frequencies of QPOs detected in different flares primarily fall within the range of 0.2--0.5\,Hz (see Table~\ref{tab:1}), and do not vary with the source luminosities. Additionally, the occurrence of QPOs is highly transient and irregular. These properties pose a challenge for the KFM \citep{1987van}. This model suggests that QPOs are generated when a portion of the radiative region is blocked by the accretion disc, thus a transient QPO is not expected, and it predicts a positive relation between the QPO frequency and luminosity.


By comparing the QPO properties of RX J0440.9+4431 with other XRPs, \cite{2023Malacaria} argued that when the source enters the super-critical accretion, QPOs are fed through an accretion disc whose inner region viscosity is unstable to mass accretion rate and temperature variations. Then, assuming the BFM \citep{1985Alpar}, the magnetic field of RX J0440.9+4431 is estimated. However, the QPO centroid frequency in RX J0440.9+4431 is unrelated to the source intensity. This contrasts with the predictions of the BFM, which suggest a variation in QPO frequency with the systematic increase in mass accretion rate, as observed in cases like XTE J1858+034 \citep{2006Mukherjee}, 4U 1901+03 \citep{2011James}, EXO 2030+375 \citep{1989Angelini}. 

Based on the preceding analysis, a solid correlation between the flares and QPOs is evident. The QPOs only appear during the periods when the flares occur at the right wing of the pulse profile, and they manifest as the count rate modulation of the flares. Additionally, the flares originate from an increase in hard photons (Figs.~\ref{fig:1} and \ref{fig:5}), causing that the QPOs are only observed in the hard energy band (see Fig.~\ref{fig:4}). Therefore, to understand the generation of the QPOs, we need to comprehend the generation of the flares, as well as the modulation mechanism during the flares.

Considering that the flares appear at a fixed narrow pulse phase (Figs.~\ref{fig:1} and \ref{fig:2}), there are two scenarios to explain the generation of the flares. In the first scenario, a short-scale increase in material within the accretion flow at a fixed phase occurs. This additional material, when conveyed to the NS polar cap, may generate X-ray flashing. In the second scenario, the increase in the accretion flow is considered to be phase-independent, but the radiation produced by the additional material in the accretion column is highly beaming, thus the radiation emits at a narrow pulse phase.

To explain the energy dependence of the flares, a plausible explanation may be proposed that the different angles at which high and low-energy photons are emitted from the accretion column may cause only hard photons to be observed during the flares. However, the analysis of the pulse profile of RX J0440.9+4431 in \cite{2023lipp} indicates that the angles of emission for high and low-energy photons from the accretion column are the same during super-critical accretion. In addition, we do not observe a soft flare at another pulse phase. Therefore, the energy dependence of the flares suggests that the radiation generated by this additional material in the accretion column should be predominantly high-energy.

Taking a step further to consider the origins of modulation in the flare, the quasi-periodic modulation frequency displayed by the flare, ranging from 0.2 to 0.5\,Hz, significantly exceeds the timescales for photon diffusion from hotspots/accretion columns and the reprocessing of X-ray flux by the atmosphere \citep{2019Mushtukov,2023Zhang}. Therefore, the most likely origin of the modulation is the instability in the accretion flow. If this instability arises accidentally and is independent of the accretion rate, this would explain the transient nature of the QPO and its independence from luminosity. Regarding the origin of this instability, more theoretical or simulation work is needed in the future.

\subsection{Magnetic field strength}
The PDS of RX J0440.9+4431 shows typical aperiodic noise in a wide range of Fourier frequencies (see Fig.~\ref{fig:6}). As seen in Fig.~\ref{fig:7}, lower luminosity corresponds to a larger magnetosphere, resulting in a smaller break frequency. The blue points and blue dashed dotted line in the figure represent the results studied by \cite{2012Tsygankov} based on observations from the 2010 outburst. However, the model curve fitted with only four data points does not adequately capture the data points obtained in a broader luminosity range during our current study. We then re-fit the model, with the solid black line and dashed black line representing fitting curves that include and exclude the blue points, respectively. The difference between the two fittings is not significant. Across four orders of magnitude in luminosity, the break frequencies seem to align well with $\propto L_{\mathrm{x}}^{3 / 7}$.

Considering the parameters $\Lambda$ as 0.5, $m$ as 1.4, and $R_6$ as 1.2, with the break frequency equal to the Keplerian frequency at the inner disc radius \citep{2009Revnivtsev}, we determine the magnetic field strength of RX J0440.9+4431 to be ($7.4\pm{0.4}) \times 10^{13}$\,G. This value is consistent with the magnetic field strength derived by \cite{2023Salganik} through the pulsar spin-up rates and the transition of the pulsar to the supercritical accretion regime \citep{2015Mushtukov}. However, it is larger than the magnetic field strength obtained by \cite{2012Tsygankov}, who fitted the relationship between luminosity and break frequency using only four data points, as well as the magnetic field strength acquired through the possible cyclotron absorption feature around 30\,keV. But in further studies, \cite{2013Ferrigno} and \cite{2023Salganik} attributed the deficiency of photons at 30\,keV to an inadequate modeling of the continuum emission. During the 2022--2023 outburst of RX J0440.9+4431, the \textit{NuSTAR} observations \citep{2023Salganik} did not detect the presence of a Cyclotron Resonant Scattering Feature (CRSF). Additionally, the break frequency could be expected to correspond to the dynamo frequency $v_{d}$ at the innermost radius, representing the local initial time-scale of variability \citep{2004King,2019Mushtukov}. If a coefficient $k_{d}$ connects the two frequencies as $v_d=v_k / k_d$, we take the dynamo frequency corresponding to the factor $k_d = 2$, referring to both \cite{2014Doroshenko} and \cite{1979Ghosh}. In this scenario, the magnetic field strength $B$ is ($3.3\pm{0.2}) \times 10^{13}$\,G. By considering $k_d = 10$ as the lower limit for the break frequency \citep{2004King}, the corresponding lower limit for the magnetic field strength is ($5.0\pm{0.3}) \times 10^{12}$\,G. In summary, RX J0440.9+4431 shows a relatively strong magnetic field of around $10^{13}$\,G, corresponding to a CRSF energy of $\sim$120\,keV if the CRSF originates near the surface of NS. The high energy of the CRSF makes detection challenging, which could explain why we did not observe any cyclotron absorption features in the spectra. Additionally, even if we consider the uncertainties in the parameters $\Lambda$, $m$, $R_6$, it does not change the fact that this source has a strong magnetic field.


\section{Conclusion}
\label{sec:5}

The broadband and high-statistics observations of RX J0440.9+4431 during its 2022--2023 giant outburst by \textit{Insight}-HXMT and \textit{NICER} allow us to conduct a detailed study of aperiodic variability in this source. Above the critical luminosity, occasional flares with timescales on the order of seconds are observed at the peak of the pulse profile, primarily caused by an increase in hard photons. We have identified five flares in the \textit{Insight}-HXMT observations that exhibit QPOs. Consistent with the findings reported by \cite{2023Malacaria}, these QPOs show phase dependence. The centroid frequencies of the QPOs range from 0.2 to 0.5\,Hz and are independent of the source intensity. The rms of the QPOs significantly increases with energy in the 10--60\,keV range, and signals are still detectable above 75\,keV. The characteristics of QPOs in RX J0440.9+4431 challenge commonly used BFM \citep{1985Alpar} and KFM \citep{1987van}. We consider that the QPO in hard flares is associated with instabilities in the accretion flow. The phase dependence of the flares can be explained by an increase in material ingested by the accretion column from a fixed phase, primarily generating hard photons. Alternatively, there is another possibility that the increase in accretion rate, independent of the phase, results in highly beamed hard photons produced within the accretion column.


Additionally, by analyzing \textit{NICER} data with longer GTIs, we investigate the relationship between the break frequencies of the noise power spectra and luminosity in RX J0440.9+4431. We find that the break frequencies align well with $\propto L_{\mathrm{x}}^{3 / 7}$. Considering that the break frequency equals the Keplerian frequency/dynamo frequency at the inner radius of the accretion disc, we estimate the magnetic field $B$ of RX J0440.9+4431 to be \textasciitilde$10^{13}$\,G, and provide a lower limit of ($5.0\pm{0.3}) \times 10^{12}$\,G.

\section*{Acknowledgements}

This study made use of data and software resources from the High Energy Astrophysics Science Archive Research Center (HEASARC), generously provided by NASA's Goddard Space Flight Center. Additionally, we leveraged the resources of the \textit{Insight}-HXMT mission, which receives support from the China National Space Administration (CNSA) and the Chinese Academy of Sciences (CAS). Financial support for this work is derived from the National Key R\&D Program of China (2021YFA0718500). We also express our gratitude for funding received from the National Natural Science Foundation of China (NSFC) under grant numbers 12122306, 12333007, U2038102, U2031205, U2038104, 12173103, 12041303, as well as the CAS Pioneer Hundred Talent Program Y8291130K2, the Strategic Priority Research Program of the Chinese Academy of Sciences under grant No. XDB0550300 and the Scientific and Technological Innovation Project of IHEP Y7515570U1.

\section*{Data Availability}

\textit{Insight}-HXMT data is accessible through the following website: \url{http://hxmtweb.ihep.ac.cn/}. Additionally, the \textit{NICER} data utilized in this study can be publicly accessed via the High Energy Astrophysics Science Archive Research Centre (HEASARC) website at \url{https://heasarc.gsfc.nasa.gov/cgi-bin/W3Browse/w3browse.pl}.



\bibliographystyle{mnras}
\bibliography{example} 

\begin{thebibliography}{}
\makeatletter
\relax
\def\mn@urlcharsother{\let\do\@makeother \do\$\do\&\do\#\do\^\do\_\do\%\do\~}
\def\mn@doi{\begingroup\mn@urlcharsother \@ifnextchar [ {\mn@doi@} {\mn@doi@[]}}
\def\mn@doi@[#1]#2{\def\@tempa{#1}\ifx\@tempa\@empty \href {http://dx.doi.org/#2} {doi:#2}\else \href {http://dx.doi.org/#2} {#1}\fi \endgroup}
\def\mn@eprint#1#2{\mn@eprint@#1:#2::\@nil}
\def\mn@eprint@arXiv#1{\href {http://arxiv.org/abs/#1} {{\tt arXiv:#1}}}
\def\mn@eprint@dblp#1{\href {http://dblp.uni-trier.de/rec/bibtex/#1.xml} {dblp:#1}}
\def\mn@eprint@#1:#2:#3:#4\@nil{\def\@tempa {#1}\def\@tempb {#2}\def\@tempc {#3}\ifx \@tempc \@empty \let \@tempc \@tempb \let \@tempb \@tempa \fi \ifx \@tempb \@empty \def\@tempb {arXiv}\fi \@ifundefined {mn@eprint@\@tempb}{\@tempb:\@tempc}{\expandafter \expandafter \csname mn@eprint@\@tempb\endcsname \expandafter{\@tempc}}}

\bibitem[\protect\citeauthoryear{{Alpar} \& {Shaham}}{{Alpar} \& {Shaham}}{1985}]{1985Alpar}
{Alpar} M.~A.,  {Shaham} J.,  1985, \mn@doi [\nat] {10.1038/316239a0}, \href {https://ui.adsabs.harvard.edu/abs/1985Natur.316..239A} {316, 239}

\bibitem[\protect\citeauthoryear{{Angelini}, {Stella}  \& {Parmar}}{{Angelini} et~al.}{1989}]{1989Angelini}
{Angelini} L.,  {Stella} L.,   {Parmar} A.~N.,  1989, \mn@doi [\apj] {10.1086/168070}, \href {https://ui.adsabs.harvard.edu/abs/1989ApJ...346..906A} {346, 906}

\bibitem[\protect\citeauthoryear{Bachetti et~al.,}{Bachetti et~al.}{2022}]{matteo2022}
Bachetti M.,  et~al., 2022, StingraySoftware/stingray: Version 1.0, \mn@doi{10.5281/zenodo.6394742}, \url {https://doi.org/10.5281/zenodo.6394742}

\bibitem[\protect\citeauthoryear{{Bailer-Jones}, {Rybizki}, {Fouesneau}, {Demleitner}  \& {Andrae}}{{Bailer-Jones} et~al.}{2021}]{2021Bailer-Jones}
{Bailer-Jones} C.~A.~L.,  {Rybizki} J.,  {Fouesneau} M.,  {Demleitner} M.,   {Andrae} R.,  2021, VizieR Online Data Catalog, \href {https://ui.adsabs.harvard.edu/abs/2021yCat.1352....0B} {p. I/352}

\bibitem[\protect\citeauthoryear{{Belloni}, {Psaltis}  \& {van der Klis}}{{Belloni} et~al.}{2002}]{2002Belloni}
{Belloni} T.,  {Psaltis} D.,   {van der Klis} M.,  2002, \mn@doi [\apj] {10.1086/340290}, \href {https://ui.adsabs.harvard.edu/abs/2002ApJ...572..392B} {572, 392}

\bibitem[\protect\citeauthoryear{{Bodaghee} et~al.,}{{Bodaghee} et~al.}{2016}]{2016Bodaghee}
{Bodaghee} A.,  et~al., 2016, \mn@doi [\apj] {10.3847/0004-637X/823/2/146}, \href {https://ui.adsabs.harvard.edu/abs/2016ApJ...823..146B} {823, 146}

\bibitem[\protect\citeauthoryear{{Cao} et~al.,}{{Cao} et~al.}{2020}]{2020SCPMACao}
{Cao} X.,  et~al., 2020, \mn@doi [Science China Physics, Mechanics, and Astronomy] {10.1007/s11433-019-1506-1}, \href {https://ui.adsabs.harvard.edu/abs/2020SCPMA..6349504C} {63, 249504}

\bibitem[\protect\citeauthoryear{{Chen} et~al.,}{{Chen} et~al.}{2020}]{2020SCPMAChen}
{Chen} Y.,  et~al., 2020, \mn@doi [Science China Physics, Mechanics, and Astronomy] {10.1007/s11433-019-1469-5}, \href {https://ui.adsabs.harvard.edu/abs/2020SCPMA..6349505C} {63, 249505}

\bibitem[\protect\citeauthoryear{{Churazov}, {Gilfanov}  \& {Revnivtsev}}{{Churazov} et~al.}{2001}]{2001Churazov}
{Churazov} E.,  {Gilfanov} M.,   {Revnivtsev} M.,  2001, \mn@doi [\mnras] {10.1046/j.1365-8711.2001.04056.x}, \href {https://ui.adsabs.harvard.edu/abs/2001MNRAS.321..759C} {321, 759}

\bibitem[\protect\citeauthoryear{{Coley} et~al.,}{{Coley} et~al.}{2023}]{2023ATelColey}
{Coley} J.~B.,  et~al., 2023, The Astronomer's Telegram, \href {https://ui.adsabs.harvard.edu/abs/2023ATel15907....1C} {15907, 1}

\bibitem[\protect\citeauthoryear{{Devasia}, {James}, {Paul}  \& {Indulekha}}{{Devasia} et~al.}{2011}]{2011Devasia}
{Devasia} J.,  {James} M.,  {Paul} B.,   {Indulekha} K.,  2011, \mn@doi [\mnras] {10.1111/j.1365-2966.2011.18407.x}, \href {https://ui.adsabs.harvard.edu/abs/2011MNRAS.414.1023D} {414, 1023}

\bibitem[\protect\citeauthoryear{{Doroshenko}, {Santangelo}, {Doroshenko}, {Caballero}, {Tsygankov}  \& {Rothschild}}{{Doroshenko} et~al.}{2014}]{2014Doroshenko}
{Doroshenko} V.,  {Santangelo} A.,  {Doroshenko} R.,  {Caballero} I.,  {Tsygankov} S.,   {Rothschild} R.,  2014, in European Physical Journal Web of Conferences. p. 06009, \mn@doi{10.1051/epjconf/20136406009}

\bibitem[\protect\citeauthoryear{{Doroshenko} et~al.,}{{Doroshenko} et~al.}{2020}]{2020Doroshenko}
{Doroshenko} V.,  et~al., 2020, \mn@doi [\mnras] {10.1093/mnras/stz2879}, \href {https://ui.adsabs.harvard.edu/abs/2020MNRAS.491.1857D} {491, 1857}

\bibitem[\protect\citeauthoryear{{Doroshenko} et~al.,}{{Doroshenko} et~al.}{2023}]{2023Doroshenko}
{Doroshenko} V.,  et~al., 2023, \mn@doi [arXiv e-prints] {10.48550/arXiv.2306.02116}, \href {https://ui.adsabs.harvard.edu/abs/2023arXiv230602116D} {p. arXiv:2306.02116}

\bibitem[\protect\citeauthoryear{{Ferrigno}, {Farinelli}, {Bozzo}, {Pottschmidt}, {Klochkov}  \& {Kretschmar}}{{Ferrigno} et~al.}{2013}]{2013Ferrigno}
{Ferrigno} C.,  {Farinelli} R.,  {Bozzo} E.,  {Pottschmidt} K.,  {Klochkov} D.,   {Kretschmar} P.,  2013, \mn@doi [\aap] {10.1051/0004-6361/201321053}, \href {https://ui.adsabs.harvard.edu/abs/2013A&A...553A.103F} {553, A103}

\bibitem[\protect\citeauthoryear{{Frank}, {King}  \& {Raine}}{{Frank} et~al.}{2002}]{2002Frank}
{Frank} J.,  {King} A.,   {Raine} D.~J.,  2002, {Accretion Power in Astrophysics: Third Edition}

\bibitem[\protect\citeauthoryear{{Gendreau} et~al.,}{{Gendreau} et~al.}{2016}]{2016SPIE}
{Gendreau} K.~C.,  et~al., 2016, in {den Herder} J.-W.~A.,  {Takahashi} T.,   {Bautz} M.,  eds,  Society of Photo-Optical Instrumentation Engineers (SPIE) Conference Series Vol. 9905, Space Telescopes and Instrumentation 2016: Ultraviolet to Gamma Ray. p. 99051H, \mn@doi{10.1117/12.2231304}

\bibitem[\protect\citeauthoryear{{Ghosh} \& {Lamb}}{{Ghosh} \& {Lamb}}{1979}]{1979Ghosh}
{Ghosh} P.,  {Lamb} F.~K.,  1979, \mn@doi [\apj] {10.1086/157285}, \href {https://ui.adsabs.harvard.edu/abs/1979ApJ...232..259G} {232, 259}

\bibitem[\protect\citeauthoryear{{Ingram} \& {Done}}{{Ingram} \& {Done}}{2012}]{2012Ingram}
{Ingram} A.,  {Done} C.,  2012, \mn@doi [\mnras] {10.1111/j.1365-2966.2011.19885.x}, \href {https://ui.adsabs.harvard.edu/abs/2012MNRAS.419.2369I} {419, 2369}

\bibitem[\protect\citeauthoryear{{James}, {Paul}, {Devasia}  \& {Indulekha}}{{James} et~al.}{2010}]{2010James}
{James} M.,  {Paul} B.,  {Devasia} J.,   {Indulekha} K.,  2010, \mn@doi [\mnras] {10.1111/j.1365-2966.2010.16880.x}, \href {https://ui.adsabs.harvard.edu/abs/2010MNRAS.407..285J} {407, 285}

\bibitem[\protect\citeauthoryear{{James}, {Paul}, {Devasia}  \& {Indulekha}}{{James} et~al.}{2011a}]{2011Jamesb}
{James} M.,  {Paul} B.,  {Devasia} J.,   {Indulekha} K.,  2011a, \mn@doi [\mnras] {10.1111/j.1365-2966.2010.17543.x}, \href {https://ui.adsabs.harvard.edu/abs/2011MNRAS.410.1489J} {410, 1489}

\bibitem[\protect\citeauthoryear{{James}, {Paul}, {Devasia}  \& {Indulekha}}{{James} et~al.}{2011b}]{2011James}
{James} M.,  {Paul} B.,  {Devasia} J.,   {Indulekha} K.,  2011b, \mn@doi [\mnras] {10.1111/j.1365-2966.2010.17543.x}, \href {https://ui.adsabs.harvard.edu/abs/2011MNRAS.410.1489J} {410, 1489}

\bibitem[\protect\citeauthoryear{{King}, {Pringle}, {West}  \& {Livio}}{{King} et~al.}{2004}]{2004King}
{King} A.~R.,  {Pringle} J.~E.,  {West} R.~G.,   {Livio} M.,  2004, \mn@doi [\mnras] {10.1111/j.1365-2966.2004.07322.x}, \href {https://ui.adsabs.harvard.edu/abs/2004MNRAS.348..111K} {348, 111}

\bibitem[\protect\citeauthoryear{{Lai}}{{Lai}}{2014}]{2014Lai}
{Lai} D.,  2014, in European Physical Journal Web of Conferences. p. 01001 (\mn@eprint {arXiv} {1402.1903}), \mn@doi{10.1051/epjconf/20136401001}

\bibitem[\protect\citeauthoryear{{Li}, {Wang}  \& {Zhao}}{{Li} et~al.}{2012}]{2012Li}
{Li} J.,  {Wang} W.,   {Zhao} Y.,  2012, \mn@doi [\mnras] {10.1111/j.1365-2966.2012.21096.x}, \href {https://ui.adsabs.harvard.edu/abs/2012MNRAS.423.2854L} {423, 2854}

\bibitem[\protect\citeauthoryear{{Li} et~al.,}{{Li} et~al.}{2023}]{2023lipp}
{Li} P.~P.,  et~al., 2023, \mn@doi [\mnras] {10.1093/mnras/stad2956}, \href {https://ui.adsabs.harvard.edu/abs/2023MNRAS.526.3637L} {526, 3637}

\bibitem[\protect\citeauthoryear{{Liu} et~al.,}{{Liu} et~al.}{2020}]{2020SCPMALiu}
{Liu} C.,  et~al., 2020, \mn@doi [Science China Physics, Mechanics, and Astronomy] {10.1007/s11433-019-1486-x}, \href {https://ui.adsabs.harvard.edu/abs/2020SCPMA..6349503L} {63, 249503}

\bibitem[\protect\citeauthoryear{{Lyubarskii}}{{Lyubarskii}}{1997}]{1997Lyubarskii}
{Lyubarskii} Y.~E.,  1997, \mn@doi [\mnras] {10.1093/mnras/292.3.679}, \href {https://ui.adsabs.harvard.edu/abs/1997MNRAS.292..679L} {292, 679}

\bibitem[\protect\citeauthoryear{{Ma} et~al.,}{{Ma} et~al.}{2022}]{2022Ma}
{Ma} R.,  et~al., 2022, \mn@doi [\mnras] {10.1093/mnras/stac2768}, \href {https://ui.adsabs.harvard.edu/abs/2022MNRAS.517.1988M} {517, 1988}

\bibitem[\protect\citeauthoryear{{Malacaria}, {Huppenkothen}, {Roberts}, {Ducci}, {Bozzo}, {Jenke}, {Wilson-Hodge}  \& {Falanga}}{{Malacaria} et~al.}{2023}]{2023Malacaria}
{Malacaria} C.,  {Huppenkothen} D.,  {Roberts} O.~J.,  {Ducci} L.,  {Bozzo} E.,  {Jenke} P.,  {Wilson-Hodge} C.~A.,   {Falanga} M.,  2023, \mn@doi [arXiv e-prints] {10.48550/arXiv.2310.16498}, \href {https://ui.adsabs.harvard.edu/abs/2023arXiv231016498M} {p. arXiv:2310.16498}

\bibitem[\protect\citeauthoryear{{Mandal} et~al.,}{{Mandal} et~al.}{2023}]{2023Mandal}
{Mandal} M.,  et~al., 2023, \mn@doi [arXiv e-prints] {10.48550/arXiv.2306.08083}, \href {https://ui.adsabs.harvard.edu/abs/2023arXiv230608083M} {p. arXiv:2306.08083}

\bibitem[\protect\citeauthoryear{{McHardy}, {Papadakis}, {Uttley}, {Mason}  \& {Page}}{{McHardy} et~al.}{2004}]{2004McHardy}
{McHardy} I.~M.,  {Papadakis} I.~E.,  {Uttley} P.,  {Mason} K.~O.,   {Page} M.~J.,  2004, \mn@doi [Nuclear Physics B Proceedings Supplements] {10.1016/j.nuclphysbps.2004.04.015}, \href {https://ui.adsabs.harvard.edu/abs/2004NuPhS.132..122M} {132, 122}

\bibitem[\protect\citeauthoryear{{M{\"o}nkk{\"o}nen}, {Tsygankov}, {Mushtukov}, {Doroshenko}, {Suleimanov}  \& {Poutanen}}{{M{\"o}nkk{\"o}nen} et~al.}{2022}]{2022MM}
{M{\"o}nkk{\"o}nen} J.,  {Tsygankov} S.~S.,  {Mushtukov} A.~A.,  {Doroshenko} V.,  {Suleimanov} V.~F.,   {Poutanen} J.,  2022, \mn@doi [\mnras] {10.1093/mnras/stac1828}, \href {https://ui.adsabs.harvard.edu/abs/2022MNRAS.515..571M} {515, 571}

\bibitem[\protect\citeauthoryear{{Motch}, {Haberl}, {Dennerl}, {Pakull}  \& {Janot-Pacheco}}{{Motch} et~al.}{1997}]{1997Motch}
{Motch} C.,  {Haberl} F.,  {Dennerl} K.,  {Pakull} M.,   {Janot-Pacheco} E.,  1997, \mn@doi [\aap] {10.48550/arXiv.astro-ph/9611122}, \href {https://ui.adsabs.harvard.edu/abs/1997A&A...323..853M} {323, 853}

\bibitem[\protect\citeauthoryear{{Mukherjee}, {Bapna}, {Raichur}, {Paul}  \& {Jaaffrey}}{{Mukherjee} et~al.}{2006}]{2006Mukherjee}
{Mukherjee} U.,  {Bapna} S.,  {Raichur} H.,  {Paul} B.,   {Jaaffrey} S.~N.~A.,  2006, \mn@doi [Journal of Astrophysics and Astronomy] {10.1007/BF02702648}, \href {https://ui.adsabs.harvard.edu/abs/2006JApA...27...25M} {27, 25}

\bibitem[\protect\citeauthoryear{{Mushtukov} \& {Tsygankov}}{{Mushtukov} \& {Tsygankov}}{2022}]{2022Mushtukov}
{Mushtukov} A.,  {Tsygankov} S.,  2022, \mn@doi [arXiv e-prints] {10.48550/arXiv.2204.14185}, \href {https://ui.adsabs.harvard.edu/abs/2022arXiv220414185M} {p. arXiv:2204.14185}

\bibitem[\protect\citeauthoryear{{Mushtukov}, {Suleimanov}, {Tsygankov}  \& {Poutanen}}{{Mushtukov} et~al.}{2015}]{2015Mushtukov}
{Mushtukov} A.~A.,  {Suleimanov} V.~F.,  {Tsygankov} S.~S.,   {Poutanen} J.,  2015, \mn@doi [\mnras] {10.1093/mnras/stu2484}, \href {https://ui.adsabs.harvard.edu/abs/2015MNRAS.447.1847M} {447, 1847}

\bibitem[\protect\citeauthoryear{{Mushtukov}, {Ingram}  \& {van der Klis}}{{Mushtukov} et~al.}{2018}]{2018Mushtukovp}
{Mushtukov} A.~A.,  {Ingram} A.,   {van der Klis} M.,  2018, \mn@doi [\mnras] {10.1093/mnras/stx2872}, \href {https://ui.adsabs.harvard.edu/abs/2018MNRAS.474.2259M} {474, 2259}

\bibitem[\protect\citeauthoryear{{Mushtukov}, {Lipunova}, {Ingram}, {Tsygankov}, {M{\"o}nkk{\"o}nen}  \& {van der Klis}}{{Mushtukov} et~al.}{2019}]{2019Mushtukov}
{Mushtukov} A.~A.,  {Lipunova} G.~V.,  {Ingram} A.,  {Tsygankov} S.~S.,  {M{\"o}nkk{\"o}nen} J.,   {van der Klis} M.,  2019, \mn@doi [\mnras] {10.1093/mnras/stz948}, \href {https://ui.adsabs.harvard.edu/abs/2019MNRAS.486.4061M} {486, 4061}

\bibitem[\protect\citeauthoryear{{Nakajima} et~al.,}{{Nakajima} et~al.}{2022}]{2022Nakajima}
{Nakajima} M.,  et~al., 2022, The Astronomer's Telegram, \href {https://ui.adsabs.harvard.edu/abs/2022ATel15835....1N} {15835, 1}

\bibitem[\protect\citeauthoryear{{Pal} et~al.,}{{Pal} et~al.}{2023}]{2023ATelpal}
{Pal} S.,  et~al., 2023, The Astronomer's Telegram, \href {https://ui.adsabs.harvard.edu/abs/2023ATel15868....1P} {15868, 1}

\bibitem[\protect\citeauthoryear{{Qu}, {Zhang}, {Song}  \& {Falanga}}{{Qu} et~al.}{2005}]{2005Qu}
{Qu} J.~L.,  {Zhang} S.,  {Song} L.~M.,   {Falanga} M.,  2005, \mn@doi [\apjl] {10.1086/444350}, \href {https://ui.adsabs.harvard.edu/abs/2005ApJ...629L..33Q} {629, L33}

\bibitem[\protect\citeauthoryear{{Raichur} \& {Paul}}{{Raichur} \& {Paul}}{2008}]{2008Raichur}
{Raichur} H.,  {Paul} B.,  2008, \mn@doi [\apj] {10.1086/591037}, \href {https://ui.adsabs.harvard.edu/abs/2008ApJ...685.1109R} {685, 1109}

\bibitem[\protect\citeauthoryear{{Reig}}{{Reig}}{2008}]{2008Reig}
{Reig} P.,  2008, \mn@doi [\aap] {10.1051/0004-6361:200810021}, \href {https://ui.adsabs.harvard.edu/abs/2008A&A...489..725R} {489, 725}

\bibitem[\protect\citeauthoryear{{Reig}, {Negueruela}, {Fabregat}, {Chato}  \& {Coe}}{{Reig} et~al.}{2005}]{2005Reig}
{Reig} P.,  {Negueruela} I.,  {Fabregat} J.,  {Chato} R.,   {Coe} M.~J.,  2005, \mn@doi [\aap] {10.1051/0004-6361:20053124}, \href {https://ui.adsabs.harvard.edu/abs/2005A&A...440.1079R} {440, 1079}

\bibitem[\protect\citeauthoryear{{Revnivtsev}, {Churazov}, {Postnov}  \& {Tsygankov}}{{Revnivtsev} et~al.}{2009}]{2009Revnivtsev}
{Revnivtsev} M.,  {Churazov} E.,  {Postnov} K.,   {Tsygankov} S.,  2009, \mn@doi [\aap] {10.1051/0004-6361/200912317}, \href {https://ui.adsabs.harvard.edu/abs/2009A&A...507.1211R} {507, 1211}

\bibitem[\protect\citeauthoryear{{Salganik}, {Tsygankov}, {Doroshenko}, {Molkov}, {Lutovinov}, {Mushtukov}  \& {Poutanen}}{{Salganik} et~al.}{2023}]{2023Salganik}
{Salganik} A.,  {Tsygankov} S.~S.,  {Doroshenko} V.,  {Molkov} S.~V.,  {Lutovinov} A.~A.,  {Mushtukov} A.~A.,   {Poutanen} J.,  2023, \mn@doi [arXiv e-prints] {10.48550/arXiv.2304.14881}, \href {https://ui.adsabs.harvard.edu/abs/2023arXiv230414881S} {p. arXiv:2304.14881}

\bibitem[\protect\citeauthoryear{{Stiele} \& {Yu}}{{Stiele} \& {Yu}}{2015}]{2015Stiele}
{Stiele} H.,  {Yu} W.,  2015, \mn@doi [\mnras] {10.1093/mnras/stv1530}, \href {https://ui.adsabs.harvard.edu/abs/2015MNRAS.452.3666S} {452, 3666}

\bibitem[\protect\citeauthoryear{{Tian} \& {Zou}}{{Tian} \& {Zou}}{2014}]{2014Tian}
{Tian} J.,  {Zou} Y.-C.,  2014, \mn@doi [arXiv e-prints] {10.48550/arXiv.1401.2525}, \href {https://ui.adsabs.harvard.edu/abs/2014arXiv1401.2525T} {p. arXiv:1401.2525}

\bibitem[\protect\citeauthoryear{{Tsygankov}, {Krivonos}  \& {Lutovinov}}{{Tsygankov} et~al.}{2012}]{2012Tsygankov}
{Tsygankov} S.~S.,  {Krivonos} R.~A.,   {Lutovinov} A.~A.,  2012, \mn@doi [\mnras] {10.1111/j.1365-2966.2012.20475.x}, \href {https://ui.adsabs.harvard.edu/abs/2012MNRAS.421.2407T} {421, 2407}

\bibitem[\protect\citeauthoryear{{Wang}}{{Wang}}{2016}]{2016Wang}
{Wang} J.,  2016, \mn@doi [International Journal of Astronomy and Astrophysics] {10.4236/ijaa.2016.61006}, \href {https://ui.adsabs.harvard.edu/abs/2016IJAA....6...82W} {6, 82}

\bibitem[\protect\citeauthoryear{{Zhang} et~al.,}{{Zhang} et~al.}{2020}]{zhang2020}
{Zhang} S.-N.,  et~al., 2020, \mn@doi [Science China Physics, Mechanics, and Astronomy] {10.1007/s11433-019-1432-6}, \href {https://ui.adsabs.harvard.edu/abs/2020SCPMA..6349502Z} {63, 249502}

\bibitem[\protect\citeauthoryear{{Zhang}, {Blaes}  \& {Jiang}}{{Zhang} et~al.}{2023}]{2023Zhang}
{Zhang} L.,  {Blaes} O.,   {Jiang} Y.-F.,  2023, \mn@doi [\mnras] {10.1093/mnras/stad063}, \href {https://ui.adsabs.harvard.edu/abs/2023MNRAS.520.1421Z} {520, 1421}

\bibitem[\protect\citeauthoryear{{van der Klis}, {Stella}, {White}, {Jansen}  \& {Parmar}}{{van der Klis} et~al.}{1987}]{1987van}
{van der Klis} M.,  {Stella} L.,  {White} N.,  {Jansen} F.,   {Parmar} A.~N.,  1987, \mn@doi [\apj] {10.1086/165210}, \href {https://ui.adsabs.harvard.edu/abs/1987ApJ...316..411V} {316, 411}

\makeatother
\end{thebibliography}





\bsp	
\label{lastpage}
\end{document}